\documentclass[runningheads,11pt]{llncs}
\usepackage{amsmath}
\usepackage{graphicx}
\def\calP{\mathcal{P}}

\def\calC{\mathcal{C}}

\newtheorem{observation}{Observation}
\begin{document}

\title{Algorithms for Connectivity Maintenance and Barrier Coverage
on a Closed Cycle
%\thanks{This research was supported in part by NSF under Grant CCF-1317143.}
}
\author{
Shimin Li\inst{1}
% \and
% Zhongjiang Yan\inst{2}
% \and
% Jingru Zhang\inst{3}
}
\institute{
Department of Computer Science\\
Winona State University, Winona, MN 55987, USA\\
\email{shimin.li@winona.edu}% \and
% School  of  Electronics  and  Information\\
% Northwestern  Polytechnical University, Xi'an, Shaanxi, P.R.China\\
% \email{zhjyan@nwpu.edu.cn}\and
% Department of Computer Science\\
% University  of  Texas-Rio Grande Valley, Edinburg, TX 78539, USA\\
% \email{jingru.zhang@utrgv.edu}
}

\maketitle

\begin{abstract}
The problem of maintaining connectivity of a wireless network on a closed cycle is studied in this paper.
In the initial input, we have $n$ points located on a closed cycle.
The points can move along the cycle, and
if the distance between two points is at most a given value $r$, 
we say these two points are connected.
The goal of the problem is to move the points along the cycle 
such that any adjacent pair of points is directly connected--i.e., 
there exist two paths between them in opposite directions along the cycle--while
minimizing the maximum movement over all points.
This problem is motivated by applications in mobile wireless networks, including sensors, 
vehicles, and satellites operating on closed orbits.
It is also applicable to barrier or border coverage problems, 
where sensors are deployed along a closed boundary and 
coverage is achieved through repositioning along the cycle.
We present a linear time optimal algorithm for this problem.
Then we refine the algorithm to obtain a lexicographically optimal 
solution without increasing the time complexity.
\end{abstract}

\keywords{min-max, wireless network, connectivity, barrier coverage, lexicographic optimization, linear time}

\section{Introduction}\label{intr}
The maintenance of connectivity of wireless networks built with mobile devices is critical
for the availability and reliability of the whole system.
We consider the problem of maintaining the connectivity of the wireless network on a 
closed cycle. 
All the mobile devices are represented by points located on a closed cycle initially.
The points can only move along the cycle.
If the distance between two points is not greater than a distance $r$ ($r>0$), then we say
the two points are connected by an edge, and thus can communicate with each other directly.
Two points are connected if there exists an edge or a path between them.
% If there exist two different paths between any pair of points,
% then we say these two points are 2-connected.
On the closed cycle, if
the distance between any pair of adjacent points is at most $r$, 
then there exist two paths in the opposite directions along the cycle.
The goal of the problem is to move the points along the cycle such that 
% any point is connected to its adjacent points on the cycle
the distance between any adjacent pair of points is at most $r$,
and the maximum movement over all points is minimized.
We present a linear time algorithm for this problem, and then refine the algorithm 
to obtain a lexicographically optimal solution without increasing the time complexity.

This problem is motivated by the connectivity maintenance
of ad hoc networks~\cite{YankConnectivity12}.
In a vehicular ad hoc network, each vehicle is a node in the ad hoc network
constructed by all the vehicles on the road.
If the distance of two vehicles is equal to or less than the wireless
communication radius $r$, then these two vehicles can communicate with each other
through wireless connection directly.
To guarantee the transmission of information between vehicles on accidents or
road conditions,
we would like to guarantee two paths between every pair of vehicles
in two different directions so that all the vehicles can
send and receive the information with redundancy.
The satellites around the orbit can be modeled as points on a closed cycle as well, 
and the connectivity of the satellite network can be maintained by moving the satellites along the orbit.
The algorithms also apply to the barrier or border coverage problem~\cite{Li2019A}
where the sensors are deployed on a closed cycle (border to be covered) and
the coverage of the whole border is achieved by moving the sensors along the border.
The coverage of the closed border is guaranteed if the distance between 
any pair of adjacent points is at most $2r$ where $r$ is the sensor's coverage radius.
Note that for the coverage problem, the distance between any pair of adjacent points is at most $2r$,
not $r$ as in the connectivity problem, because the point on the border is covered if 
its distance to one of the two sensors in opposite directions is at most $r$.
This difference does not affect the correctness and and efficiency of our algorithms.
For the convenience and consistence of presentation, 
we call this problem the {\em connectivity maintenance on a closed cycle problem} 
and use $r$ as the maximum distance between any pair of adjacent points in 
the rest of this paper.

\subsection{Previous Work}
The problem arises from the mobile wireless network connectivity problem~\cite{DiazOn08}
where the vertices or sensors move in random directions
% there is an edge connecting two vertices 
% if their distance is no more than a given value,
to maintain a connected graph.
D\'iaz et al. provide an analytical model for the connectivity of dynamic random geometric
graphs which is used for mobile wireless networks.
In the model, the vehicles move in random directions, and if the distance between a pair
of mobiles is not greater than a given value, there exists an edge connecting them~\cite{DiazOn08}.
A similar connectivity model for wireless ad hoc networks with communication constraints is
also presented in~\cite{Balister2010Pe,Krzywdzinski2011Ge}.
Das et al. consider some min-max movement problems of sensors in path network in plane~\cite{Das2020Op}.
The paths are edges of a graph where the sensors are located and the sensors can move in the 
2D plane to maintain the connectivity of the network.
For the case where the sensors are on a path (not a cycle), 
the problem is solved in linear time~\cite{Das2020Op,Li2019A}.
The case where the sensors are on a cycle is solved in $O(n^3)$ time 
by solving a group of simultaneous equations that 
can compute the displacements of the sensors~\cite{Das2020Op}.
For the line barrier case with the sensors 
that are initially in the 2D plane, an $O(n^2\log n)$ time algorithm is presented
by Li et al.~\cite{LI2019106841}, and then improved to $O(n^2)$ by Yao et al.~\cite{YAO2023109717}.
These are different from our problem because we require the sensors to move only along the cycle, 
while in~\cite{Das2020Op,LI2019106841,YAO2023109717} the sensors can move in the plane.
In the situations of barrier or border coverage and satellite network connectivity, 
the sensors and satellites can only move along the cycle.
Chen et al. also consider min-max movement problems of barrier coverage 
on a line or a cycle and present $O(n\log n)$ algorithms for the general case where
the sensors have arbitrary sensing ranges~\cite{Chen2013Al}.
It can be improved to $O(n)$ time for the special case where all the sensors have the same 
sensing range~\cite{Chen2013Al}, but it is not lexicographically optimal.
To our best knowledge, the lexicographically optimal algorithm to the min-max movement problem 
on a closed cycle running in linear time has not been presented before.

Some related problems on connectivity of networks are also studied in the literature.
The relationship between the node degree and $k$-connectivity of wireless multihop network
is researched by Bettstetter~\cite{Bettstetter2002On}.
Biconnectivity of wireless networks is also considered with the application of
connected mobile robots~\cite{DasA09}.
In general, $k$-connectivity problem with unreliable network links is considered and
solved by Zhao~\cite{Zhao2014Mi}.
The coverage problems of wireless sensor network attracted quite a lot of attention from
researchers and in some situations 
it is also necessary to maintain the connectivity of the wireless
network~\cite{Ghosh2008Co,Li2009Co,Liang2014A,Miorandi2005Co,Norman2011Co}.
Specifically, the probability of a static wireless ad hoc network being biconnected is
presented by Tian et al.~\cite{tian2008critical}.
Cheng et al. solve the mobile ad hoc network connectivity problem by deploying relay nodes
to the network~\cite{ChengWi14}.

\subsection{Problem Definition}
Given a distance $r>0$ and a set $\calP=\{p_1, p_2, \ldots, p_n\}$ of $n$ points on a closed cycle
$\calC$ in the input.
The points along $\calC$ are sorted in the clockwise direction without loss of generality.
Denote by $|\calC|$ the length of the cycle.
We assume that $|\calC|\leq nr$ since otherwise, there is no solution to the problem.
Define the coordinate of point $p_1$ as $x_1=0$, and
then the coordinate of each point $p_i$ is the distance from $p_1$ to $p_i$
along $\calC$ in the clockwise direction for $1<i\leq n$.
Let $c(x_i, x_j)$ be the distance between two coordinates $x_i$ and $x_j$ along
$\calC$ in the clockwise direction.
In detail, we have the following equation for any two coordinates $x_i$ and $x_j$ on $\calC$,
\begin{equation}
\label{eq:10}
c(x_i, x_j)=
\begin{cases}
x_j-x_i           & \text{if } x_i\leq x_j \\
|\calC|-(x_i-x_j) & \text{if } x_i>x_j.
\end{cases}
\end{equation}
Obviously, if $c(x_i, x_{i+1})\leq r$ for $1\leq i<n$ and $c(x_n, x_1)\leq r$,
then all the adjacent points in $\calP$ are directly connected on $\calC$.
The goal of this problem is to move the $n$ points along $\calC$ so that 
the maximum distance between any adjacent pair of points is at most $r$,
while minimizing the maximum distance that any point is moved.

\subsection{Our Approaches}
We apply an approach similar with that in~\cite{Li2025Al} to solve this problem.
The approach is based on the {\em order preserving property} of the problem.
It is obvious that the order preserving property holds in this problem,
because for any pair of points out of order in the optimal solution,
we can always swap their positions without increasing the maximum moving distance.
The main idea of our algorithm is to move the points along the cycle
in a greedy way to keep the connectivity of processed points.

Our algorithm starts with the first point and adds the following points
to the connected network one by one on the cycle
in the clockwise direction.
Each point is only moved in the counterclockwise direction if it is not connected 
to the previous point in the network.
Once we obtained a connected network of all the points on the cycle, 
we shift all the points clockwise by half of the maximum moving distance 
to obtain an optimal solution to the problem.
In detail, for each newly added point, if the distance between 
it and the previous point is greater than $r$, then we move it counterclockwise 
so that its distance to the previous point is exactly $r$;
if the distance is not greater than $r$, then we keep it at its original position.
During the process of adding and moving points,
we maintain the connectivity of the network and the maximum moving distance $d_{\max}$
of points (only in the counterclockwise direction) is minimized.
After processing all the points on the cycle,
we check whether the last point and the first point are connected.
If they are connected, 
then we move all the points $\frac{d_{\max}}{2}$ in the clockwise direction to obtain an optimal solution to
the problem. 

It might be the case that the last point and the first point are not connected
after all the points were processed.
In this situation, we continue the same process of adding points and moving them 
(only in the counterclockwise direction) to maintain the connectivity of the network.
We start from the first point and move it so the distance between the first point 
and the last point is exactly $r$.
Then we continue to add and move points in the same way as in the first round of moving points
until we meet a point that does not need to be moved in these two rounds.
Then we move all the points $\frac{d_{\max}}{2}$ in the clockwise direction to obtain an optimal solution to
the problem.
Before we present the details of our algorithm, let us first show the following observations.

\section{Observations}
By the definition of $c(x_i, x_j)$, in the equation~\eqref{eq:10}, 
we have the following observations on the distance between any pair of points on $\calC$.
\begin{observation}\label{arc_sum}
    $c(x_i, x_j)=\sum_{k=i}^{j-1}c(x_k, x_{k+1})$ for any $1\leq i<j\leq n$.
\end{observation}
This observation is true since we can divide the arc between $x_i$ and $x_j$ into 
the arcs between adjacent points $x_k$ and $x_{k+1}$ for $i\leq k<j$.
Similarly, we have the following observation on the distance between points 
where $i>j$.
\begin{observation}\label{arc_sum2}
    $c(x_i, x_j)=c(x_i, x_n)+c(x_n, x_1)+c(x_1, x_j)$ for any $1\leq j<i\leq n$.
\end{observation}
It is true because we can divide the arc to three parts:
the arc between $x_i$ and $x_n$, the arc between $x_n$ and $x_1$, and the arc between $x_1$ and $x_j$.
Note that we do not need to change the order of points on $\calC$ to obtain an optimal solution,
so we always assume that the points keep the same order in the following sections.

For the order of points does not change in the optimal solution, we 
can determine the longest distance between any pair of points in a connected network.
Let $D(x_i, x_j)$ be the maximum distance between any pair of points $x_i$ and $x_j$, 
then we have the following observation,
\begin{observation}\label{min_connected_distance}
    $D(x_i, x_j)=((n+(j-i))\mod n)\cdot r$.
\end{observation}
The observation holds when $i<j$ because the number of arcs between $x_i$ and $x_j$ 
is $j-i$ and the distance between any pair of adjacent points is at most $r$ in the optimal solution.
It also holds when $i>j$ because the number of arcs between $x_i$ and $x_j$ is 
$(n-i)+1+(j-1)=n+j-i$, which is $(n+(j-i))\mod n$.

Let us define the function $\Delta(i,j)$ for any pair of points $p_i$ and $p_j$ as follows,
\begin{equation}\label{eq:delta}
\Delta(i,j)=c(x_i,x_j)-D(x_i,x_j).
\end{equation}
Here $c(x_i,x_j)$ is the distance between $x_i$ and $x_j$ in the clockwise direction along $\calC$.
In words, $\Delta(i,j)$ is the difference between the distance of $x_i$ and $x_j$ in the original input
and the distance of $x_i$ and $x_j$ in a solution to the problem.

Denote by $d_{opt}$ the minimum moving distance of points in the optimal solution.
We have the following observation on $d_{opt}$.
\begin{observation}\label{lower_bound_observation}
    $d_{opt}\geq\max_{1\leq i,j\leq n}\frac{\Delta(i,j)}{2}$.
\end{observation}
% \begin{proof}
The observation holds because the sum of the moving distance of $p_i$ and $p_j$ 
is at least $\Delta(i,j)$ to maintain the connectivity of $p_i$ and $p_j$ in the clockwise direction.
Thus the minimum moving distance of the pair of points $p_i$ and $p_j$ is 
at least $\frac{\Delta(i,j)}{2}$.
In general, this applies to any pair of points $p_i$ and $p_j$ for $1\leq i,j\leq n$.
% Therefore, we have the observation holds.
% \qed
% \end{proof}

By the above observations, a straightforward approach to solve the problem 
is to compute $\frac{\Delta(i,j)}{2}$ for all pairs of points $p_i$ and $p_j$,
and then find the maximum value.
Since there are $O(n^2)$ pairs of points, the time complexity of this approach is $O(n^2)$.
The difficult part of the problem is to find the pair of points 
$p_i$ and $p_j$ such that $\frac{\Delta(i,j)}{2}$ is maximized in linear time.
Now let us present our algorithm to obtain an optimal 
solution in linear time.

\section{Algorithm Description}
\label{algorithm_description}
For the clear presentation of our algorithm,
we separate our algorithm into two stages: the one-direction moving stage and 
the adjusting stage.
In the one-direction moving stage, we move all the points only in the 
counterclockwise direction along $\calC$
to build a connected network.
This stage is similar to the algorithm in~\cite{Li2025Al} for the cycle case.
We continue to move the first point if it is not connected to the last point after
the first round of moving all the points.
The second round continues until we meet a point that does not need to be moved in these two rounds.
In the adjusting stage, we move all the points clockwise by half of the maximum moving distance
in the above two rounds
to obtain an optimal solution.
The adjusting stage is straightforward, 
so we focus on the one-direction moving stage in the remainder of this section.

\subsection{The One-direction Moving Stage}
\label{onedirection}
In the one-direction moving stage of our algorithm,
all the points will only be moved in the counterclockwise direction along $\calC$.
A greedy approach is applied to add and move the points in $\calP$ one by one
to maintain a connected network of processed points on $\calC$.
Let us present the details of our algorithm in this stage.

\subsubsection{The First Round}
\label{round1}
Our algorithm starts with the first point $p_1$ and adds the following points
in the clockwise direction on $\calC$.
Denote by $x'_i$ the new position of point $p_i$ after the movement.
Initially, the point $p_1$ is connected itself, so it is added to the 
network without any movement, i.e., $x'_1=x_1=0$.
It is clear that the maximum moving distance of points is $d_{\max}=0$ initially.

From now on, suppose we have just processed the point $p_{i}$ on $\calC$,
and its new position is $x'_{i}$ after the movement.
Now we are considering the next point $p_{i+1}$.
There are two different cases based on the distance between $x'_{i}$ and $x_{i+1}$.

\begin{enumerate}
\item
Case $c(x'_i, x_{i+1})\leq r$.
In this case, we do not need to move $p_{i+1}$ to connect it to the network.
Add $p_{i+1}$ to the network without moving it, i.e., let $x'_{i+1}=x_{i+1}$.
Obviously, the maximum moving distance $d_{\max}$ does not change in this case.

\item
Case $c(x'_i, x_{i+1})>r$.
Point $p_{i+1}$ needs to be moved in the counterclockwise direction to maintain
the connectivity of points in the network.
Add $p_{i+1}$ to the network and move it to $x'_{i+1}=x'_{i}+r$.
Then update the maximum moving distance $d_{\max}$ 
by the following equation 
$$d_{\max}=\max\left\{d_{\max}, c(x'_{i}, x_{i+1})-r\right\}.$$
\end{enumerate}

Once the point $p_{i+1}$ is processed, we continue to process the next point
$p_{i+2}$ in the same way until all the $n$ points are processed.

\subsubsection{The Second Round}
\label{round2}
After all the $n$ points are processed in the first round,
we check whether the distance between the last point $p_n$ and the first point $p_1$ 
is less than or equal to $r$.
If $c(x'_n, x'_1)\leq r$, then the first stage is done.
Otherwise, we need to start the second round of adding and moving points
to maintain the connectivity of the network on $\calC$.

Denote by $x''_i$ the new position of point $p_i$ after the movement
in the second round.
Initially, we move the first point $p_1$ 
to $x''_1=(x'_n+r) \mod |\calC|$ to connect it to the last point $p_n$.
Then update the maximum moving distance $d_{\max}$ by setting
$d_{\max}=\max\left\{d_{\max}, c(x'_n, x'_1)-r\right\}$.
Note that $c(x'_n, x'_1)=|\calC|-(x'_n-x'_1)$ if $x'_n>x'_1$ by the equation~\eqref{eq:10}.
In general, we process the following points in the same way as in the first round,
until we meet a point that does not need to be moved in these two rounds.
That is, we stop the second round when we meet a point $p_k$ such that
$x''_{k}=x'_{k}=x_{k}$.

The following lemma guarantees the existence of such a point on $\calC$.

\begin{lemma}\label{existence_of_fixed_point}
    There exists a point $p_k$ on $\calC$ such that
    $x''_{k}=x'_{k}=x_{k}$ after the above two rounds of moving operations.
\end{lemma}
\begin{proof}
    Note that there must be a point that is not moved in the first round 
    because the point $p_1$ is not moved at the beginning.
    Let $p_k$ be the last point on $\calC$ such that
    $x'_{k}=x_{k}$ after the first round of moving points.
    In details, we have $x'_i<x_i$ for $k<i\leq n$ after the first round of moving points.
    Now let us prove the lemma by contradiction.
    
    Assume the lemma is not true, which means that $p_k$ must be moved 
    again in the second round.
    Then we must have $x''_{i}<x'_{i}$ for $1\leq i<k$ 
    and $c(x''_{k-1},x'_k)>r$ after processing $p_{k-1}$ in the second round.
    After the first round, we must have $c(x'_{i},x'_{i+1})=r$ for $k\leq i<n$
    by the rules of moving points (we just move the minimum distance to maintain 
    the connectivity of points).
    After the second round, we have $c(x''_{n},x''_{1})=r$ and 
    $c(x''_{i},x''_{i+1})=r$ for $1\leq i<k-1$ for the same reason.
    Therefore, the sum of the distances between all adjacent pairs of points 
    must satisfy $|\calC|=c(x'_k,x'_n)+c(x'_n,x''_1)+c(x''_1,x''_{k-1})+c(x''_{k-1},x'_k)$.
    We have $c(x'_k,x'_n)=(n-k)r$, $c(x'_n,x''_1)=r$, and $c(x''_1,x''_{k-1})=(k-2)r$ 
    when processing the point $p_k$ in the second round of moving points.
    That means we have the following equation holding.
    \begin{equation}\label{eq:cycle_length}
    |\calC|=(n-k+1+k-2)r+c(x''_{k-1},x'_k)=(n-1)r+c(x''_{k-1},x'_k) 
    \end{equation}
    
    Note that we have $c(x''_{k-1},x'_k)>r$ based on our assumption that $p_k$ 
    needs to be moved in the second round.
    That means the following inequality holds,
    \begin{equation}
    |\calC|>(n-1)r+r=n\cdot r 
    \end{equation}
    which contradicts our assumption that $|\calC|\leq nr$ in the problem definition.
    Clearly, if $|\calC|>nr$, then we do not have enough points to maintain the 
    connectivity between each pair of adjacent points on $\calC$.

    Therefore, there exists a point $p_k$ on $\calC$ such that
    $x''_{k}=x'_{k}=x_{k}$ after the above two rounds of moving operations.
\qed
\end{proof}

By the above Lemma~\ref{existence_of_fixed_point}, our algorithm will stop some time in the second round.
That is the end of the one-direction moving stage of our algorithm.
Once the one-direction moving stage is done,
we move all the points clockwise by $\frac{d_{\max}}{2}$ 
to obtain an optimal solution to the problem.

\section{Correctness and Optimality of the Algorithm}
\label{correctness_analysis}
Now let us show the correctness of our algorithm.
The following theorem guarantees that the outputs of our algorithm 
is correct.

\begin{theorem}\label{correctness_theorem}
    The distance between any adjacent pair of points on $\calC$ 
    (including $p_1$ and $p_n$) is 
    at most $r$ after the processing of our algorithm.
\end{theorem}
\begin{proof}
    Obviously, the distance between any adjacent pair of points 
    does not change in the adjusting stage of our algorithm 
    where we move all the points clockwise by 
    the same distance, $\frac{d_{\max}}{2}$.
    Thus, it is sufficient to prove that all the points are connected
    after the one-direction moving stage.

    During the one-direction moving stage,
    we always maintain the connectivity of processed points on $\calC$
    in the first round because the algorithm moves the next point
    if it is not connected to the previous point.
    Therefore, after the first round is done,
    the distance between any adjacent pair of points $p_i$ and $p_{i+1}$ 
    for $1\leq i<n$ is at most $r$.
    
    Now let us consider the second round of moving points.
    The first point processed in the second round is $p_1$.
    We move $p_1$ if it is not directly connected to the last point $p_n$ after
    the first round, then check whether $c(x''_1, x'_{2})>r$.
    In general, only the distance between $p_i$ and $p_{i+1}$ 
    might be greater than $r$ after processing the point $p_i$ in the second round.
    If the distance between $p_i$ and $p_{i+1}$ is at most $r$,
    then the algorithm does not move $p_{i+1}$.
    That means we met the point $p_k$ such that $x''_{k}=x'_{k}$ where $k=i+1$.
    The algorithm stops here with all the pairs of adjacent points directly connected on $\calC$.

    On the other hand, if we have $c(x''_{i}, x'_{i+1})>r$,
    the algorithm moves $p_{i+1}$ to connect it to $p_i$
    in the second round.
    The process continues until we meet a point $p_k$ such that
    the distance between $p_k$ and $p_{k-1}$ is at most $r$.
    Then we do not move $p_k$ in the second round and the algorithm stops 
    with all pairs of adjacent points directly connected on $\calC$.

    By Lemma~\ref{existence_of_fixed_point}, there exists a point 
    $p_k$ on $\calC$
    such that $x''_{k}=x'_{k}=x_{k}$ after the above two rounds of moving operations.
    Therefore, the distance between any adjacent pair of points on $\calC$ 
    (including $p_1$ and $p_n$) is 
    at most $r$ after the processing of our algorithm.
\qed
\end{proof}

Therefore, the output of our algorithm is a solution to the problem.
By Lemma~\ref{existence_of_fixed_point}, we know that there must be a point 
that is not moved in the first stage.
Define the consecutive points that moved in the first stage as a 
{\em moving sequence}.
There is a point that is not moved right before (in the counterclockwise direction)
the moving sequence on $\calC$.
Let $p_{s}$ be a point that is not moved in the first stage, then we have $x''_s=x'_s=x_s$
(or $x'_s=x_s$ if $p_s$ is not involved in the second round).
For the points in the moving sequence right after $p_s$, 
we have the following lemma on their moving distances.
\begin{lemma}\label{dmax_delta}
    The point $p_j$ is moved 
    $\Delta(s,j)$ for any point $p_j$ in the moving sequence right after $p_s$.
\end{lemma}
\begin{proof}
    Let us prove the case where $1\leq s<j\leq n$ first. Then we can 
    extend the result to the general cases where the moving sequence crosses
    the points $p_n$ and $p_1$.

    For the case where $1\leq s<j\leq n$, we 
    can prove it by induction.
    If there is only one point $p_{s+1}$ in the moving sequence,
    we have $x'_{s+1}=x'_{s}+r$ by our algorithm.
    Then the moving distance of $p_{s+1}$ is $c(x_{s}, x_{s+1})-r=\Delta(s,s+1)$.
    The base case is proved.

    For the inductive step, assume that the result holds 
    for the $k$-th point in the moving sequence, 
    i.e., the moving distance of $p_{s+k}$ is $\Delta(s,s+k)$, 
    where $\Delta(s,s+k)=c(x_{s}, x_{s+k})-D(x_{s}, x_{s+k})$.

    Now consider the next point $p_{s+k+1}$ in the moving sequence.
    By Observation~\ref{arc_sum}, we have $c(x_s, x_{s+k+1})=c(x_s, x_{s+k})+c(x_{s+k}, x_{s+k+1})$.
    We have $D(x_s, x_{s+k+1})=D(x_s, x_{s+k})+r$ by Observation~\ref{min_connected_distance}.
    Based on the inductive assumption, we can determine the moving distance 
    of $p_{s+k+1}$ is $\Delta(s,s+k+1)=c(x_{s}, x_{s+k+1})-D(x_{s}, x_{s+k+1})$.

    Therefore, by induction, 
    we have the lemma holds for all points in the moving sequence right after $p_s$
    where $1\leq s<j\leq n$.

    Now let us consider the case where $s>j$, i.e., the moving sequence crosses the points $p_n$ and $p_1$.
    That means the points $p_1$, $p_2$, $\ldots$, $p_j$ are moved in the second round due 
    to $c(x'_n, x'_1)>r$ after the first round.
    By Lemma~\ref{existence_of_fixed_point}, all the points moved in the second round
    are in one moving sequence because we stop the second round when we meet a point 
    that does not need to be moved.
    For the simplicity of analysis, 
    we can add $|\calC|$ to the coordinates of points that did not move across the 
    origin of $\calC$ in the second round 
    (in other words, we extend the coordinate system along the cycle).
    Similarly, we can also add $n$ to the indices of these points.
    Then we can apply the same induction above to prove the lemma for the case where $s>j$.
    The only difference is that we need to use Observation~\ref{arc_sum2} instead of Observation~\ref{arc_sum}
    to calculate the distance between $x_s$ and $x_j$ in the original input.
\qed
\end{proof}

The above lemma leads to the following corollary.

\begin{corollary}\label{dmax_corollary}
    The maximum moving distance $d_{\max}$ of points in the first stage
    satisfies $d_{\max}=\max_{1\leq i,j\leq n}\Delta(i,j)$.
\end{corollary}
\begin{proof}
    The Observation~\ref{lower_bound_observation} and Theorem~\ref{correctness_theorem} imply that 
    \begin{equation}\label{eq:lower_bound}
        d_{\max}\geq 2\cdot d_{opt}\geq\max_{1\leq i,j\leq n}\Delta(i,j).
    \end{equation}
    On the other hand, by Lemma~\ref{dmax_delta}, we have $d_{\max}=\Delta(s,j^*)$ 
    where $p_{j^*}$ is the point that is moved $d_{\max}$ in the first stage
    and $p_s$ is the point that is not moved in the first stage before 
    $p_{j^*}$ in the clockwise direction.
    Therefore, we have     
    \begin{equation}\label{eq:upper_bound}
        d_{\max}=\Delta(s,j^*)\leq\max_{1\leq i,j\leq n}\Delta(i,j).
    \end{equation}
    By the above two inequalities \ref{eq:lower_bound} and \ref{eq:upper_bound}, 
    we conclude 
    \begin{equation}\label{eq:optimality}
        d_{\max}=\max_{1\leq i,j\leq n}\Delta(i,j).
    \end{equation}
\qed
\end{proof}

The above corollary implies that the maximum moving distance $d_{\max}$ of points in the first stage
is twice of the optimal moving distance $d_{opt}$ by Observation~\ref{lower_bound_observation}.
Thus, by moving all the points counterclockwise by $\frac{d_{\max}}{2}$ in the adjusting stage, 
we can obtain an optimal solution to the problem.
Furthermore, each point is moved at most three times in our algorithm.
For each point, the calculation related to its movement can be done in constant time,
so the time complexity of our algorithm is $O(n)$.
\begin{theorem}\label{optimality_theorem}
    Our algorithm can find an optimal solution to the connectivity maintenance on a closed cycle problem
    in $O(n)$ time.
\end{theorem}

\section{The Lexicographically Optimal Algorithm}
\label{lexicographically_optimal}
The optimal solution obtained by the above algorithm is not necessarily lexicographically optimal.
That is, some points in the optimal solution may be moved an unnecessary large distance.
In this section, we present an algorithm to find a lexicographically optimal solution to 
the problem without increasing the time complexity.
The main idea of the algorithm is to extend the algorithm for connectivity maintenance 
on a line~\cite{Li2025Al} to the cycle case.

In the process of the lexicographically optimal algorithm, we separate the points into several 
connected groups on $\calC$. For each group, we find the minimum moving distance of points 
to connect them. To differentiate it from the moving sequence in the one-direction 
moving stage of the previous algorithm, we name the group of points a {\em block}.
In each block, the adjacent points are connected to each other, 
and the maximum moving distance of points in the block is minimized.
Furthermore, all the points in a block can be treated as a whole.
Our algorithm finds the lexicographically optimal solution by building the blocks that are connected 
on $\calC$.

Before we present our algorithm, let us show some definitions and properties of the blocks.
Our algorithm maintains a list of blocks $L=\{B_1,B_2,\ldots, B_j,\ldots,B_m\}$ 
where $B_j$ is the $j$-th block on $\calC$ in the clockwise direction.
Let $l_j$ and $r_j$ be the indices of the leftmost point and the rightmost point in block $B_j$, respectively.
Define the distance between two adjacent blocks $B_j$ and $B_{j+1}$ as $c(x_{r_j}, x_{l_{j+1}})$
for $1\leq j<m$.
The maximum moving distance of points in block $B_j$ is denoted by $D_j$.
The following properties of the blocks hold during the process of our algorithm.
\begin{enumerate}
\item
    The distance between each pair of adjacent points in a block is exactly $r$.
\item
    In each block, there exists a pair of points that move toward each other 
    with the maximum distance among all points in the block ($D_j$).
\item
    For any two adjacent blocks $B_j$ and $B_{j+1}$, we have 
    $c(x_{r_j}, x_{l_{j+1}})<r$.
\end{enumerate}

By the above properties, we have the following observations.
\begin{observation}\label{block_treatment}
    All the points in a block can be treated as a whole in our algorithm.
\end{observation}
This observation follows from the first property of the blocks that the distance 
between each pair of adjacent points in a block is exactly $r$.
That distance does not change once a point is added to a block, 
so all the points in a block have the same moving distance 
when we move the block as a whole.
Thus, the point that moves the maximum distance in a direction 
always moves the largest distance in that direction.
Therefore, we just need to consider the movement of the leftmost and rightmost points 
when we move the block as a whole in our algorithm.
Once the leftmost and rightmost points in a block are determined, 
we can locate the positions of all the other points by the first property in 
linear time.

\begin{observation}\label{block_optimality}
    The maximum moving distance of points in a block is minimized.
\end{observation}
\begin{proof}
    The second property of the blocks implies that there are two points 
    in a block that move $D_j$ towards each other
    where $D_j$ is the maximum moving distance of points in block $B_j$.
    The amount of $D_j$ is determined by the distance between these two 
    points in the original input and 
    the number of points between them in the clockwise direction on $\calC$
    by the Observation~\ref{lower_bound_observation}.
    Obviously, the distance between these two points cannot be extended 
    because the distance between each pair of adjacent points in a block 
    is exactly $r$ (the largest possible distance) by the first property of the blocks.
    
    Further, regardless the direction in which the block is moved, 
    the maximum moving distance of points in the block will monotonically increase.
    In detail, when we move the block in the clockwise (or counterclockwise) direction, 
    the point that moves $D_j$ in the clockwise (or counterclockwise) direction 
    will move more than $D_j$, respectively.

    Therefore, the maximum moving distance of points in a block is minimized.
\qed
\end{proof}

By the above Observation~\ref{block_optimality},
we just need to maintain $x'_{l_j}$, $x'_{r_j}$, and $D_j$ for the block $B_j$ during the algorithm.
It is not necessary to track the moving distance of each point.
% This strategy works in both directions on $\calC$.
In the process of our algorithm, 
the blocks may be built, moved, and merged when
adding a new point to the connected network.
After processing the new added point, all the above properties of the blocks are maintained.
Now let us present our algorithm in detail.

\subsection{The First Round}
Initially, we add the point $p_1$ to the first block $B_1$ without moving it, 
i.e., $x'_{1}=x_{1}$, and set $D_1=0$ and $d_{\max}=0$.
There is one block in the list $L=\{B_1\}$ at the beginning.
Clearly, all the above properties of the blocks hold in the initial state 
of our algorithm.
Now we present the general process of our algorithm to add the next point $p_{i+1}$ 
for $1\leq i<n$.

\subsubsection{The General Process of Adding a New Point}
\label{add_next_point}
Suppose we have just processed the point $p_i$ 
and the blocks in $L=\{B_1,B_2,\ldots,B_m\}$ satisfy the above properties.
Now we are considering the next point $p_{i+1}$.
Based on the distance between $x'_{i}$ and $x_{i+1}$,
there are three different cases:

\noindent{\bf The Case $c(x'_{i}, x_{i+1})<r$:}
In this case, we do not need to move $p_{i+1}$, i.e., let $x'_{i+1}=x_{i+1}$.
Since $c(x'_{i}, x_{i+1})<r$, we add $p_{i+1}$ to a new block $B_{m+1}$.
The maximum moving distance of points in block $B_{m+1}$ is $D_{m+1}=0$.
% because we do not move $p_{i+1}$ in this case.
At last, we add $B_{m+1}$ to the end of the list $L$ and increment $m$ by one.
It is obvious that all the above properties of the blocks hold 
after processing $p_{i+1}$ in this case.

\noindent{\bf The Case $r\leq c(x'_{i}, x_{i+1})\leq r+D_m$:}
Move $p_{i+1}$ to $x'_{i}+r$ and add it to the block $B_m$.
The maximum moving distance of points in block $B_m$ does not change 
because the moving distance of $p_{i+1}$ is at most $D_m$ in this case.
All the above properties of the blocks still hold because 
only the point $p_{i+1}$ is moved a distance at most $D_m$.
% in this case.

\noindent{\bf The Case $c(x'_{i}, x_{i+1})>r+D_m$:}
In this case, the point $p_{i+1}$ needs to be moved more than $D_m$ 
to connect it to $p_i$.
We move $p_{i+1}$ to $x'_{i}+r$ and add it to the block $B_m$ first.
Then we move $B_m$
by $\frac{c(x'_{i}, x_{i+1})-r-D_m}{2}$ in the clockwise direction.
The maximum moving distance of points in block $B_m$ is updated to
$D_m+\frac{c(x'_{i}, x_{i+1})-r-D_m}{2}=\frac{D_m+c(x'_{i}, x_{i+1})-r}{2}$.
In detail, the point that moves the maximum distance in the clockwise direction 
and $p_{i+1}$ move $\frac{D_m+c(x'_{i}, x_{i+1})-r}{2}$ towards each other.
They are the new pair of points in block $B_m$ that move the maximum distance 
after the movement.
Update the maximum moving distance $d_{\max}$ to
$\max\{d_{\max}, D_m\}$ to maintain the maximum moving distance of points.
In this case, the distance between any adjacent pair of points in block $B_m$ 
is exactly $r$.
There are two points in block $B_m$ that move the maximum distance
towards each other.
After moving block $B_m$ clockwise, we need to check whether $c(x'_{r_{m-1}}, x'_{l_m})<r$.
If it is true, then the process of adding $p_{i+1}$ stops.
All the properties of the blocks hold after processing $p_{i+1}$ in this case.

On the other hand, if $c(x'_{r_{m-1}}, x'_{l_m})\geq r$,
we need to run the merging process of our algorithm to maintain the 
third property of the blocks.

\subsubsection{Merging Adjacent Blocks}
Now let us present the merging process of two adjacent blocks $B_{m-1}$ and $B_m$.
This process is triggered when the distance between 
$B_{m-1}$ and $B_m$ is greater than or equal to $r$, i.e., $c(x'_{r_{m-1}}, x'_{l_m})\geq r$,
after moving $B_m$ clockwise in the case where $c(x'_{i}, x_{i+1})>r+D_m$.
Note that the maximum moving distances of points in block $B_{m-1}$ and $B_m$ 
are $D_{m-1}$ and $D_m$, respectively, before the merging process.
Based on the values of $D_{m-1}$, $D_m$, and $c(x'_{r_{m-1}}, x'_{l_m})$, 
there are three different cases:
\begin{enumerate}
\item $c(x'_{r_{m-1}}, x'_{l_m})-r\leq D_{m-1}-D_m$ \\
In this case, we move $B_{m}$ counterclockwise such that 
$c(x'_{r_{m-1}}, x'_{l_m})=r$.
Update the final coordinate of $p_{i+1}$ according to the movement of $B_m$.
The maximum moving distance of points in $B_{m-1}$ does not change because
the maximum moving distance of points in $B_m$ is 
not greater than $D_{m-1}$.
Then we merge $B_{m}$ into $B_{m-1}$ and delete $B_{m}$ from the list $L$.
At last, the number of blocks is decremented by one, i.e., update $m$ to $m-1$.
It is easy to verify that all the properties of the blocks still hold.

\item $|D_{m}-D_{m-1}|<c(x'_{r_{m-1}}, x'_{l_m})-r$ \\
We need to move blocks $B_{m-1}$ and $B_m$ towards each other such that 
$c(x'_{r_{m-1}}, x'_{l_m})=r$ in this case.
In detail, the moving distances of $B_{m-1}$ and $B_m$ ($d_{m-1}$ and $d_m$)
are obtained by the following equations.
\begin{equation}
\label{eq:merge_case2}
\begin{cases}
d_{m-1} &= \frac{c(x'_{r_{m-1}}, x'_{l_m})-r-D_{m-1}+D_m}{2} \\
d_m     &= \frac{c(x'_{r_{m-1}}, x'_{l_m})-r+D_{m-1}-D_m}{2}
\end{cases}
\end{equation}
After the movement, we update $D_{m-1}$ to $D_{m-1}+d_{m-1}$.
It is easy to verify that we have the following equation holds after the movement.
\begin{equation}
\label{eq:merge_case2_distance}
    D_{m-1}+d_{m-1}=D_m+d_m=\frac{c(x'_{r_{m-1}}, x'_{l_m})-r+D_{m-1}+D_m}{2}
\end{equation}
The maximum moving distance $d_{\max}$ is updated to $\max\{d_{\max}, D_{m-1}\}$.
The final coordinate of $p_{i+1}$ is also updated according to the movement of $B_m$.
After the movement, we merged blocks $B_{m-1}$ and $B_m$.
If $c(x'_{r_{m-2}}, x'_{l_{m-1}})<r$ holds after the merging, 
then we stop the merging process here and 
it is easy to verify that all the properties of the blocks still hold.

If $c(x'_{r_{m-2}}, x'_{l_{m-1}})\geq r$ and
$m-1>1$ (i.e., there are two or more blocks left in the list $L$),
we need to repeat the above merging process 
for blocks $B_{m-2}$ and $B_{m-1}$.
This repetition continues until there is only one block left in the list $L$
or we have $c(x'_{r_{t-1}}, x'_{l_t})<r$ for 
the last two blocks $B_{t-1}$ and $B_t$ in the list $L$.
Then the properties of the blocks hold after the merging operations.
At last, delete the merged blocks from list $L$ and 
update the number of blocks $m$ accordingly.
% The final coordinate of $p_{i+1}$ is maintained according to the movement of 
% the merged blocks in this case.

\item $c(x'_{r_{m-1}}, x'_{l_m})-r\leq D_{m}-D_{m-1}$ \\
In this case, we move $B_{m-1}$ to $B_m$ such that
$c(x'_{r_{m-1}}, x'_{l_m})=r$.
The process is similar to the above case except that we do not 
need to move $B_m$ in the first merging process.
Thus, we omit the details of the process.
\end{enumerate}

By the properties of the blocks, we can verify that all the points are connected on $\calC$ 
after the process of adding each point.
We only need to check if $p_n$ and $p_1$ are directly connected after processing $p_n$, 
i.e. $c(x'_{n}, x'_{1})\leq r$.
If $c(x'_{n}, x'_{1})\leq r$, then the algorithm stops here.
Otherwise, we continue to run the second round of merging presented in Section~\ref{lexi_round2}.

\subsection{The Second Round of Merging}\label{lexi_round2}
At the beginning of this round of merging, 
we have $m$ ($m>1$) blocks with the properties mentioned above on $\calC$.
Due to $c(x'_{n}, x'_{1})>r$ after processing $p_n$, 
we move and merge the blocks on $\calC$ until the distance between any adjacent blocks 
is less than $r$ or there is only one block left in the list $L$.
Before the presentation of the merging process, we first show the approach to maintain 
the index and coordinate of each point on $\calC$.

For the index of each point $p_i$, we add $n$ to the index of $p_{i-1}$ if $i-1\leq 0$,
i.e., we treat $p_0$ as $p_n$ and $p_{-1}$ as $p_{n-1}$, and so on.
Similarly, if $i+1>n$, we treat $p_{i+1}$ as $p_{i+1-n}$ and $p_{i+2}$ as $p_{i+2-n}$, and so on.
For the coordinate of each point $p_i$, we add $|\calC|$ to $x'_i$ if $x'_i<0$ after the 
movement of the blocks.
Similarly, if $x'_i\geq |\calC|$, we treat $x'_i$ as $x'_i-|\calC|$ for each point $p_i$.
By the above approach, we can maintain the indices and coordinates of points in a consistent way
during the merging process in the second round of merging on $\calC$.
In the following presentation of the merging process, 
we just consider the consecutive indices and continuous coordinates of points on $\calC$ 
without repeating the above adjustment approach.
Obviously, the conversion of indices and coordinates of each point can be done in constant time, 
so it does not affect the time complexity of our algorithm.

The merging process in the second round is similar to the merging process after 
adding a new point in the first round.
Let us explain the merging of $B_m$ and $B_1$ as an example.
All the other merging processes are similar to the merging of $B_m$ and $B_1$.
Based on the values of $D_m$, $D_1$, and $c(x'_{r_m}, x'_{l_1})$, there are three different cases:
\begin{enumerate}
    \item $c(x'_{r_m}, x'_{l_1})-r\leq D_{m}-D_1$ \\
    In this case, we move $B_{1}$ counterclockwise (to $B_{m}$) such that $c(x'_{r_m}, x'_{l_1})=r$.
    Then we merge $B_{m}$ and $B_{1}$. Delete $B_{m}$ from the list $L$ and 
    decrement the number of blocks $m$.
    Because we moved block $B_1$ counterclockwise, we need to verify
    whether $c(x'_{r_1}, x'_{l_2})<r$ holds after the movement of $B_1$.
    If $c(x'_{r_1}, x'_{l_2})<r$, then we stop the algorithm and
    it is easy to verify that all the properties of the blocks still hold.
    Otherwise, we need to continue to merge blocks $B_1$ and $B_2$ recursively.

    \item $|D_{1}-D_{m}|<c(x'_{r_m}, x'_{l_1})-r$ \\
    We need to move blocks $B_{m}$ and $B_1$ towards each other such that 
    $c(x'_{r_m}, x'_{l_1})=r$ in this case.
    In detail, the moving distances of $B_{m}$ and $B_1$ ($d_{m}$ and $d_1$) 
    are obtained by the following equations.
    \begin{equation}\label{eq:block_merge_case2}
        \begin{cases}
            d_{m} &= \frac{c(x'_{r_m}, x'_{l_1})-r-D_{m}+D_1}{2} \\
            d_1   &= \frac{c(x'_{r_m}, x'_{l_1})-r+D_{m}-D_1}{2}
        \end{cases}
    \end{equation}
    It is easy to verify that we have the following equation holds after the movement.
    \begin{equation}\label{eq:block_merge_distance_case2}
        D_m+d_m=D_1+d_1=\frac{c(x'_{r_m}, x'_{l_1})-r+D_{m}+D_1}{2}
    \end{equation}
    The maximum moving distance $d_{\max}$ is updated to $\max\{d_{\max}, D_m+d_m\}$.
    We also update the largest movement distance of the new merged
    block to $D_m+d_m$.
    After the movement, we merge blocks $B_{m}$ and $B_1$.
    For the convenience of presentation, we name the merged block as $B_1$.
    If $c(x'_{r_{m-1}}, x'_{l_{1}})<r$ and $c(x'_{r_{1}}, x'_{l_2})<r$
    hold after the merging, 
    then we stop the algorithm here.
    It is easy to verify that all the properties of the blocks still hold.
    Otherwise, we continue to merge blocks recursively if the distance 
    between them is not less than $r$.
    The order of merging any two adjacent blocks does not matter because 
    the merging process moves and merges the two blocks only.
    All the other blocks are not affected by the merging process.
    In summary, we run the merging process until the distance between
    any two adjacent blocks is less than $r$ or there is only one block left in the list $L$.
    % with $c(x'_{r_1}, x'_{l_1})\leq r$.
    In either case, all the properties of the blocks hold.

\item $c(x'_{r_m}, x'_{l_1})-r\leq D_{1}-D_{m}$ \\
    In this case, we move block $B_{m}$ clockwise such that
    $c(x'_{r_{m}}, x'_{l_1})=r$.
    The process is similar to the above case except that we do not 
    need to move $B_1$ in this case.
    The details of the process are omitted here.
    After the movement, we check if $c(x'_{r_{m-1}}, x'_{l_m})<r$ holds.
    If $c(x'_{r_{m-1}}, x'_{l_m})<r$, then we stop the algorithm here.
    Otherwise, we continue to merge blocks $B_{m-1}$ and $B_1$ recursively
    after merging $B_m$ into $B_1$.
\end{enumerate}

In the process of our algorithm, we add one point at a time and maintain the properties of the blocks
in the first round.
Right after the first round, any pair of adjacent points on $\calC$ is directly connected
except for $p_n$ and $p_1$.
In the second round, the properties of the blocks are maintained after the merging operations
including the pair of points $p_n$ and $p_1$.
By the properties of the blocks, we can verify that any pair of adjacent points are directly connected on $\calC$ 
after the algorithm stops in the second round.
There are $n$ points in the input, so there are at most $n$ blocks after the first round.
Notice that
the count of blocks is reduced by one after each merging operation in the second round,
so our algorithm must stop some time.
Moreover, the process of adding each point and merging blocks can be done in constant time, 
so the time complexity of our algorithm is $O(n)$.

Finally, the lexicographical optimality of the solution obtained by our algorithm 
is guaranteed by the following lemma.
\begin{lemma}\label{lexi_optimality_lemma}
    The solution obtained by our algorithm is lexicographically optimal.
\end{lemma}
\begin{proof}
    This lemma follows from the second property of the blocks:
    In each block, there exists a pair of points that move toward each other 
    with the maximum distance among all points in the block.

    For each block, consider the pair of points that move the maximum distance towards each other.
    Any further reduction of the maximum moving distance of these two points 
    will break the connectivity of points in the block 
    because the distance between each pair of adjacent points in 
    the block is exactly $r$ (by the first property of blocks). 
    Therefore, the maximum moving distance of points in each block 
    is minimized.
    That means the solution of each block is optimal.

    This property is maintained in the process of our algorithm, 
    so we can conclude that the solution obtained by our algorithm is lexicographically optimal.
\qed
\end{proof}

The above lemma and the analysis above on the correctness and time complexity 
of our algorithm lead to the following theorem.
\begin{theorem}\label{lexi_optimality_theorem}
    Our improved algorithm can find a lexicographically optimal solution to the connectivity maintenance on a closed cycle problem in $O(n)$ time.
\end{theorem}

%%%%%%%%%%%%
\bibliographystyle{plain}
\bibliography{ref}

\end{document}